\documentclass[pre,aps,twocolumn,superscriptaddress,showpacs]{revtex4-1}
\usepackage{amsmath,bm,graphicx}
\usepackage{amssymb}
\usepackage{amsfonts}
\usepackage[usenames]{color}
\newcommand{\pd}[2]{\frac{\partial #1}{\partial #2}}
\newcommand{\dd}[2]{\frac{d #1}{d #2}}

\newcommand{\M}{\mathcal{M}}
\newcommand{\bea}{\begin{eqnarray}}
\newcommand{\eea}{\end{eqnarray}}

\begin{document}

\title{Instantaneous Gelation in Smoluchowski's Coagulation Equation Revisited}
\date{\today}

\author{Robin C. Ball}
\email{R.C.Ball@warwick.ac.uk}
\affiliation{Centre for Complexity Science, University of Warwick, Gibbet Hill Road, Coventry CV4 7AL, UK}
\affiliation{Department of Physics, University of Warwick, Gibbet Hill Road, Coventry CV4 7AL, UK}
\author{Colm Connaughton}
\email{connaughtonc@gmail.com}
\affiliation{Centre for Complexity Science, University of Warwick, Gibbet Hill Road, Coventry CV4 7AL, UK}
\affiliation{Mathematics Institute, University of Warwick, Gibbet Hill Road, Coventry CV4 7AL, UK}
\author{Thorwald H.M. Stein}
\email{t.h.stein@reading.ac.uk}
\affiliation{Department of Meteorology, University of Reading, Earley Gate, Reading, RG6 6BB, UK}
\author{Oleg Zaboronski}
\email{O.V.Zaboronski@warwick.ac.uk}
\affiliation{Mathematics Institute, University of Warwick, Gibbet Hill Road, Coventry CV4 7AL, UK}

\begin{abstract}
We study the solutions of the Smoluchowski coagulation equation with a 
regularisation term which removes clusters from the system when their mass
exceeds a specified cut-off size, $M$. We focus primarily on collision
kernels which would exhibit an instantaneous gelation transition in the absence 
of any regularisation. Numerical simulations demonstrate that for such kernels
with  monodisperse initial data, the regularised gelation time {\em decreases}
as $M$ increases, consistent with the expectation that the gelation time is
zero in the unregularised system. This decrease appears to be a logarithmically
slow function of $M$, indicating that instantaneously gelling kernels may still be 
justifiable
as physical models despite the fact that they are highly singular in the absence of
a cut-off. We also study the case when a source of 
monomers is introduced in the regularised system. In this case a stationary
state is reached. We present a complete analytic description of this 
regularised stationary
state for the model  kernel, $K(m_1,m_2)=\max\left\{m_1,m_2\right\}^{\nu}$, which
gels instantaneously when $M\to \infty$ if $\nu>1$. The stationary 
cluster size distribution decays as a stretched exponential for small cluster 
sizes and crosses over to a power law decay with exponent $\nu$ for large cluster 
sizes.  The total particle density in the stationary state slowly vanishes  as 
$(\log(M^{\nu-1}))^{-1/2}$ when $M \to \infty$. The approach to the stationary 
state is non-trivial : oscillations about the stationary state emerge from the 
interplay between the monomer injection and the cut-off, $M$, which decay very 
slowly when $M$ is large. A quantitative analysis of these oscillations is
provided for the addition model which describes the situation in which clusters 
can only grow by absorbing monomers.
\end{abstract}

\pacs{83.80.Jx}
\maketitle

\section{Introduction}

The determination of the statistical time evolution of an ensemble of particles
undergoing irreversible binary coagulation is a problem which arises in many
different contexts in the physical, chemical and biological sciences.  See 
\cite{ERN1986} for a list of applications. Of 
pre-eminent interest is the cluster size distribution, $N_m(t)$, which specifies
the average density of particles of mass, $m$, at a given time $t$. Often one
aims to derive $N_m(t)$ from a given model of the microscopic dynamics
governing cluster coagulation. This problem has been extensively studied 
for almost a century starting with the seminal work of Smoluchowski \cite{SMO1917}
who showed that, at the mean field level, the cluster size distribution
of a statistically homogeneous system evolves according to the Smoluchowski 
coagulation equation:
\begin{eqnarray}
\label{eq-SKE}
\partial_t N_m &=& \int_0^{m}dm_1 K(m_1,m-m_1) N_{m_1} N_{m-m_1}\\
 \nonumber &-&2 N_m  \int_0^{\infty}dm_1 K(m,m_1) N_{m_1} + \frac{J}{m_0}\ \delta(m-m_0).
\end{eqnarray}
Here the coagulation kernel, $K(m_1,m_2)$, is proportional to the probability
rate for a cluster of mass $m_1$ merging with a cluster of mass $m_2$.
At this level of description, it encodes all relevant information about the 
underlying micro-physics. $J$ is the rate of injection of monomers, having mass $m_0$,  into the
system. $J$ may be zero depending on the application.  In this paper, we take 
Eq.~(\ref{eq-SKE}) as our departure point. It is expected to apply when the
clusters remain well-mixed (despite aggregation), but the elucidation of the exact 
conditions under which Eq.~(\ref{eq-SKE}) is 
rigorously obtained as
the mean field limit of an underlying stochastic process can be a subtle
question. 
Mathematically inclined readers are referred to the review 
by Aldous \cite{ALD1999} for some common mathematical perspectives on this issue. 
A  physical example of a situation in which the applicability
of Eq.~(\ref{eq-SKE}) fails due to the generation of spatial correlations between 
particles by diffusive fluctuations is discussed in detail in \cite{CRZ2006}.

In many applications the microphysics is scale invariant, at least over some range 
of cluster sizes.  That is to say, the coagulation kernel is a homogeneous
function of its arguments, the degree of homogeneity of which we shall
denote by $\lambda$:
\begin{equation}
\label{eq-lambda}
K(c m_1, c m_2)= c^{\lambda}K(m_1,m_2).
\end{equation}
In many cases, such kernels result in solutions of Eq.~(\ref{eq-SKE}) which
exhibit self-similarity. This means that the cluster
size distribution asymptotically takes the form $N(m,t) \sim s(t)^a F(m/s(t))$ where
$s(t)$ is a characteristic cluster size which grows in time, $a$ is a 
dynamical scaling exponent and $\sim$ denotes the scaling limit: $t\to\infty$ and
$m \to \infty$ with $m/s(t)$ fixed.
Furthermore, it has been proven for the constant, sum and product kernels that such self-similar solutions are 
attracting for any initial data with finite
$(\lambda+1)^\mathrm{th}$ moment \cite{MP2004}.
Much work in the physics literature on the theory of coagulation, following the work of
Van Dongen and Ernst \cite{EVD1988}, has focused on the problem of determining
the properties of the scaling function $F(x)$ and the exponent $a$ from the 
scaling properties of $K(m_1,m_2)$.  An almost complete scaling theory of 
Eq.~(\ref{eq-SKE}) is now known. See
the review by Leyvraz \cite{LEY2003} for a modern discussion.

A key feature of
this scaling theory, and one which is of considerable importance for what follows,
is the fact that the scaling properties of Eq.~(\ref{eq-SKE}) are often
very sensitive to the rate of coagulation of clusters of widely
different masses. In order to parameterise this dependence in a general
way, and following the notation of \cite{LEY2003}, we introduce the scaling
exponents $\mu$ and $\nu$ which specify the asymptotic behaviour of the 
coagulation kernel in the limit where the mass of one cluster greatly
exceeds that of the other:
\begin{equation}
\label{eq-muAndnu}
K( m_1,  m_2) \sim m_1^\mu m_2^\nu\ \ m_1\!\! \ll \!\! m_2.
\end{equation}
Clearly from Eq.~(\ref{eq-lambda}), we must have $\mu+\nu=\lambda$. We shall use
the notation $\mathcal{M}_\alpha(t)$ to denote the $\alpha$-moment of the
cluster size distribution:
\begin{equation}
\mathcal{M}_\alpha(t) = \int_0^\infty m^\alpha\,N_m(t)\, d m.
\end{equation}

\begin{figure}
\begin{center}
\includegraphics[width=7.0cm]{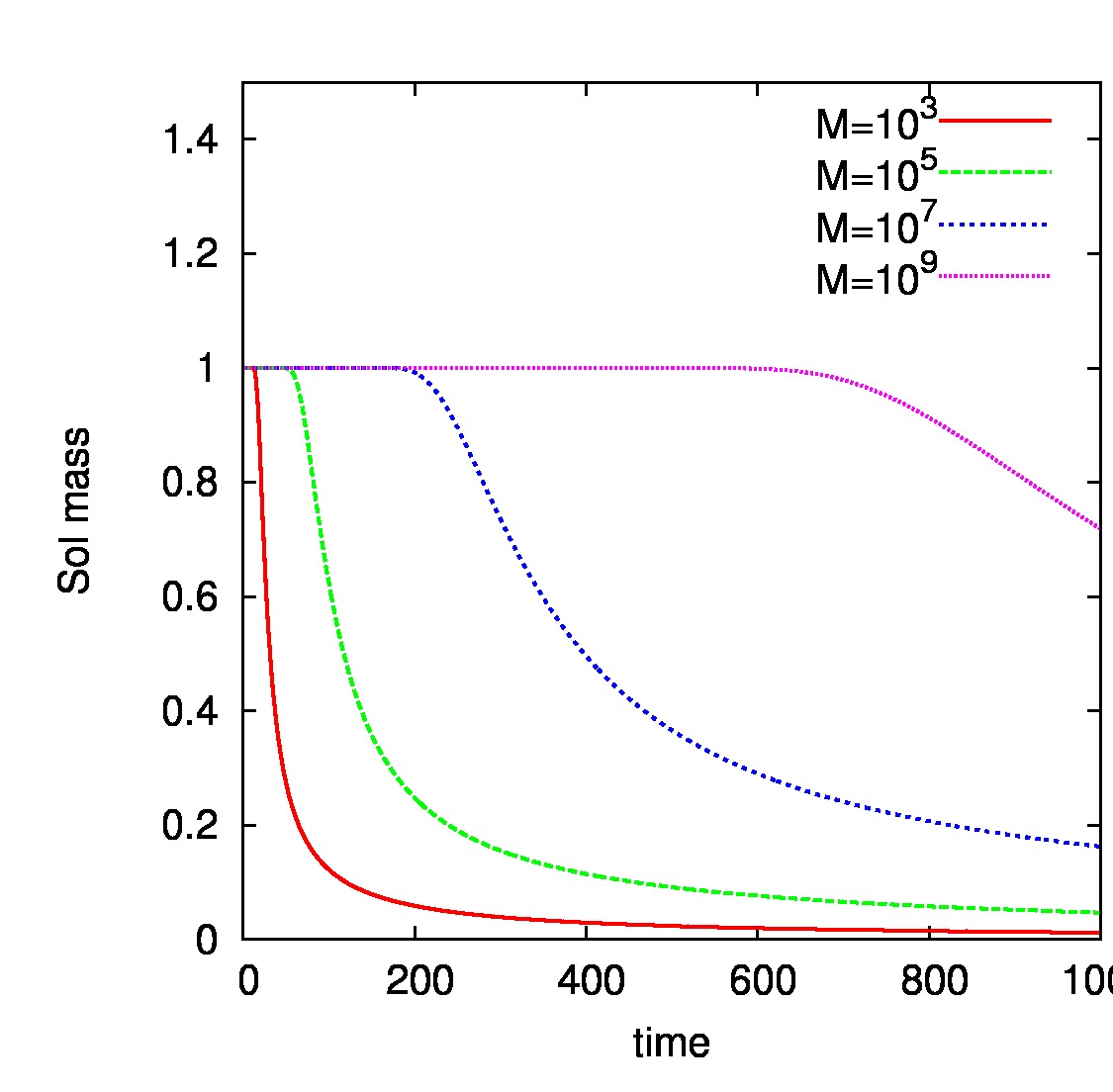}\\
\caption{\label{fig-noGelation}(Color online) Total mass density, $\M_1(t)$, for $K(m_1,m_2)=(m_1 m_2)^{1/4}$.}
\end{center}
\end{figure}

\begin{figure}
\begin{center}
\includegraphics[width=7.0cm]{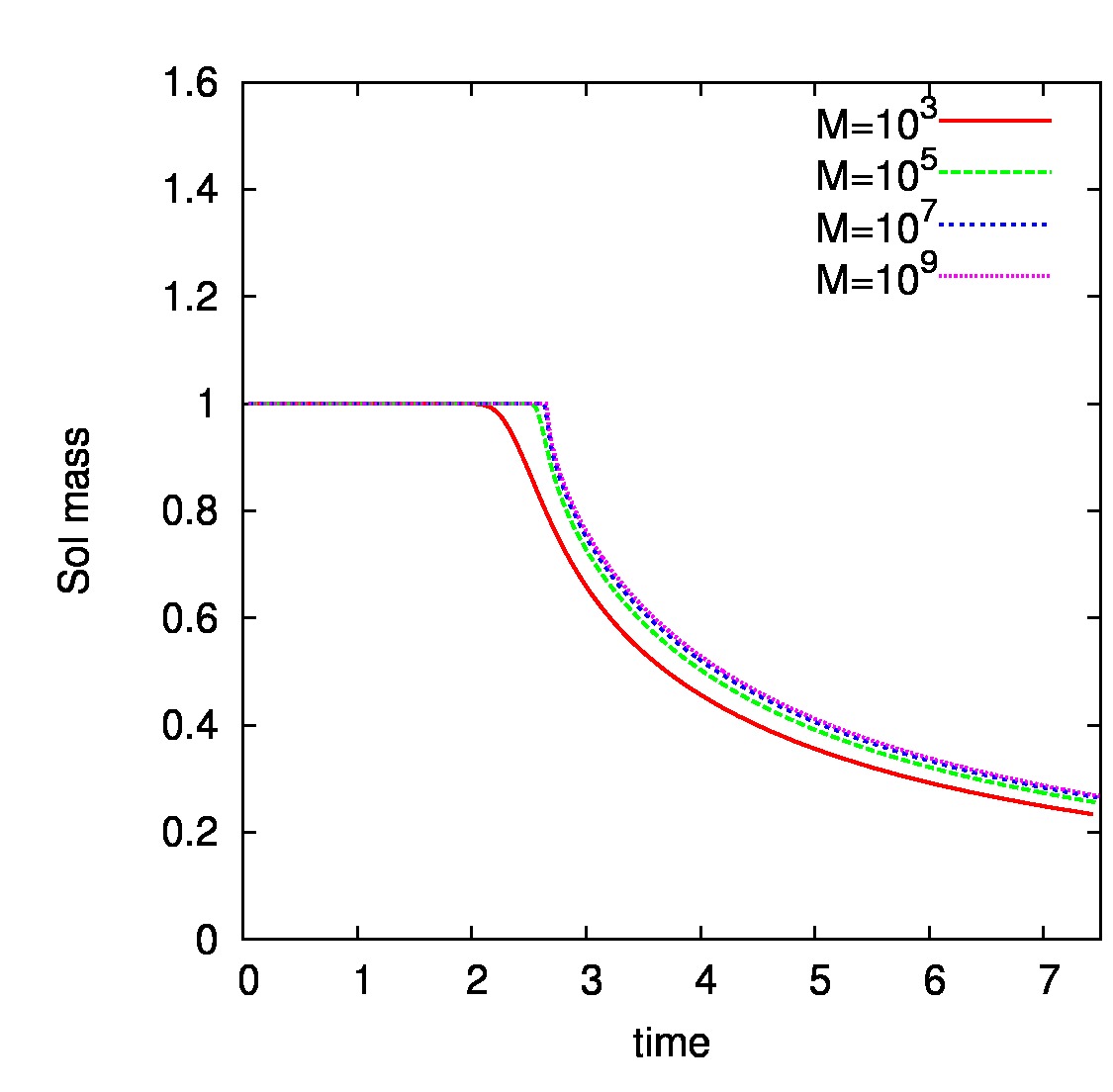}\\
\caption{\label{fig-regularGelation}(Color online) Total mass density, $\M_1(t)$, for $K(m_1,m_2)=(m_1 m_2)^{3/4}$.}
\end{center}
\end{figure}

The first moment, the total mass density, 
$\mathcal{M}_1(t)=\int_0^\infty m\,N_m(t)\,dm$, is governed by conservation of
material and is thus formally
conserved by Eq.~(\ref{eq-SKE}) when $J=0$. If $J\neq 0$ we simply have
$\mathcal{M}_1(t)=\mathcal{M}_1(0) + J\,t$. This accords with the intuition imparted by the
mass-conserving character of the individual coagulation events. This intuition is
challenged, however, by one of the more interesting phenomena to emerge from the
scaling theory of the Smoluchowski equation: the so-called gelation transition. 
Originally conjectured by Lushnikov \cite{LUS1977} and Ziff \cite{ZIF80} and put on
a firmer theoretical footing by Van Dongen and Ernst \cite{VDE1986}, gelation refers
to the fact that, for kernels having $\lambda>1$, mass-conservation can be 
spontaneously broken in finite time. That is to say, there exists a time
$t^*$, known as the gelation time such that 
\begin{displaymath}
\M_1(t) < \int_0^\infty m\,N_m(0)\,dm \ \ \mbox{$t>t^*$}.
\end{displaymath}
The characteristic cluster size, $s(t)$, typically diverges at the gelation time.
The ``missing'' mass can be interpreted as going into a cluster of infinite mass
or ``gel''. The generation of arbitrarily large clusters in finite
time might seem surprising but can be perfectly physical in some cases. One example
is polymer gelation in which clusters merge by the formation of crosslinks
and therefore do not need to move in order to coalesce. For other examples of kernels 
having $\lambda>1$, the solution of Eq.~(\ref{eq-SKE}) describes only the 
intermediate asymptotics of the
underlying physical system over some range of cluster sizes and for times less
than the gelation time. Once this intermediate asymptotic range has been
identified, there is no conceptual problem with the loss of mass conservation
in the Smoluchowski or the generation of infinite clusters. Indeed, the
gelation transition has even been observed experimentally, for example in polymer 
aggregation \cite{LNP1990}, and found to exhibit dynamics in reasonable agreement 
with the predictions of the Smoluchowski equation.

It turns out, however, that this is not the full story. Soon after the discovery of
the gelation phenomenon, it was realised by Hendriks, Ernst and Ziff \cite{HEZ1983} that
the kernel $K(m_1,m_2)=(m_1m_2)^2$ produced a series expansion for the
second moment of the cluster size distribution which seemed to have zero radius of convergence 
in time. This led the authors to suggest the possibilty of gelation occuring at time $0+$. Detailed 
study of the scaling properties of the Smoluchowski equation subsequently led to formal arguments 
\cite{VDO1987}, later made rigourous \cite{CC1992},
which show that for coagulation kernels having exponent $\nu>1$, the gelation time 
is actually zero. This surprising phenomenon is referred to as {\em instantaneous} 
gelation. Even more surprisingly, the gelation process
can be {\em complete} in the sense that $\M_1(t)=0$ for $t>0$. See \cite{JEO1999} for
a mathematical discussion of this point. A situation in which
all mass vanishes from the system in time $0+$ clearly cannot describe even the
intermediate asymptotics of any physical coagulation problem. For applications
such as the coagulation of polymers or colloidal aggregates, the fact that the
available surface area of an aggregate cannot grow faster than its volume means
that the exponent $\nu$ cannot be greater than 1. If the only applications of
the Smoluchowski equation came from polymer science, the phenomenon of
instantaneous gelation would be regarded as a mathematical pathology which need 
not concern physicists. Yet there {\em are}  models for which it can plausibly be 
argued that the exponent
$\nu$ is greater than 1. One such example is the astrophysical phenomenon 
of gravitational clustering \cite{SW1978,KON01} which is thought to play a role in 
determining the large scale matter distribution of the universe. A
second important example is that of differential 
sedimentation of water droplets falling at their terminal velocity 
\cite{KC2002,HNS2008}, one of the processes responsible for the observed droplet
size distribution in clouds \cite{FFS2002,FSV2006}. Furthermore, there are even
proposed heuristic solutions of Eq.~(\ref{eq-SKE}) in the literature for
such models \cite{KON01,HNS2008} which seem  reasonably supported by numerical
simulations. The question of how this is possible, given the known 
mathematical results on  instantaneous gelation discussed above, is the principal 
topic of this paper. It should be clear from the outset that the presence of an instantaneous
gelation transition for a particular kernel indicates that the underlying physical model
omits some process which becomes important for short timescales or for large masses. For
example, in the case of droplet coalescence in clouds, the Stokes assumption for the
terminal velocity of a droplet is ultimately responsible for the exponent $\nu$ 
exceeding 1. This assumption ceases to be valid for sufficiently large droplets 
(see \cite{KC2002} and the references therein). 

\section{Instantaneous Gelation in the Regularised System}

The gelation phenomenon in particle systems is best
understood by considering a regularisation of the system and studying the
behaviour as this regularisation is removed. Two natural 
regularisations can be found in the literature. One approach is to consider the 
stochastic dynamics of a finite number of particles as has been done by
Lushnikov \cite{LUS2005} for the product kernel. Another approach is to 
introduce a mass cut-off, $M$,  into the Smoluchowski equation.
This has been done by Filbet and Lauren{\c c}ot \cite{FL2004}. These are different
regularisations but both show singular behaviour as the regularisation is 
removed when $\lambda>1$. 

In this paper, we regularise by the latter method. As in \cite{FL2004}, the 
cut-off is introduced in such a way that clusters having mass $m>M$
are removed from the system:
\begin{eqnarray}
\label{eq-regSKE}
\partial_t N_m &=& \frac{1}{2}\,\int_1^{m}dm_1 K(m_1,m-m_1) N_{m_1} N_{m-m_1}\\
 \nonumber &-& N_m  \int_1^{M-m}dm_1 K(m,m_1) N_{m_1} + J\ \delta(m-1)\\
\nonumber &-& D_M\left[ N_m(t)\right]
\end{eqnarray}
where
\begin{equation}
\label{eq-dissipation}
D_M\left[ N_m(t)\right] = N_m \int_{M-m}^{M} dm_1\, K(m,m_1)\, N_{m_1}
\end{equation}
describes the removal of clusters having $m>M$.  This regularisation explicitly 
breaks mass conservation. Furthermore, we shall implicitly measure all masses
in terms of the monomer mass from now on so that the lower cut-off, $m_0$, is set
equal to 1. Clearly this is not the only way in which the problem can be regularised.
In particular, by omitting $D_M\left[ N_m(t)\right] $ from Eq.~(\ref{eq-regSKE}) one
would obtain an explicitly conservative regularisation. One would expect the behaviour
to be completely different in this case. The differences between  conservative and 
non-conservative regularisations for a related aggregation--fragmentation model arising
in the kinetics of waves are explored in \cite{CON2009}. The fact that the gelation
transition involves a loss of mass from the system suggests that the non-conservative
regularisation is more natural. While we implement this regularisation as a hard cut-off, $M$, 
we would expect the behaviour described below to be qualitatively the same 
for smoother regularisations provided that the non-conservative property is retained.

Figs. \ref{fig-noGelation} and \ref{fig-regularGelation} show the mass contained in 
this regularised system as a function of time for a sequence of values of 
$M$ for the non-gelling kernel $K(m_1,m_2)=(m_1 m_2)^{1/4}$ 
(Fig.~\ref{fig-noGelation}) and  for the gelling kernel $K(m_1,m_2)=(m_1 m_2)^{3/4}$
(Fig.~\ref{fig-regularGelation}).
These were obtained by numerical solution of Eq.~(\ref{eq-regSKE}) with
monodisperse initial data. 

All numerical results in this paper were obtained using the 
algorithm developed by Lee in \cite{LEE2000,LEE2001}. This algorithm involves coarse-graining
the masses into bins whose width increases exponentially with mass and then using the 
Smoluchowski equation to approximate the mass transferred per timestep as a result of  the 
aggregation of clusters in each pair of bins. The approximation explicitly enforces mass
conservation. Full details of
the coarse-graining and formulae for the computation of mass transfer are provided in 
\cite{LEE2000}. In a slight modification of Lee's algorithm, the regularisation described by 
Eq.~(\ref{eq-dissipation}) is implemented by introducing an infinite width bin, 
$\left[M, \infty\right]$, which accumulates any mass transferred to clusters larger than $M$. 
We also modified the timestepping procedure, replacing Lee's original explicit timestepping
method with an adaptive implicit trapezoidal rule. This is essential because the
Smoluchowski equation becomes increasingly stiff for larger values of $\lambda$ to the
extent that the results presented here could not be obtained with explicit timestepping. 
The implicit trapezoidal rule requires the solution of a set of nonlinear equations at each 
timestep which was done using the GSL \cite{GSL} implementation of the Rosenbrock algorithm 
\cite{ROS1960}. Explicit formulae and further discussion of these modifications of Lee's
algorithm can be found in \cite{CON2009}.

We see that for the non-gelling system, mass conservation 
is restored in the limit $M \to \infty$ whereas for the gelling system,
it is not. This latter situation will be recognisable to readers familiar with the
theory of turbulence where it is widely believed that energy conservation is
not restored when the limit of zero viscosity is taken in the Navier-Stokes
equations \cite{FRI1995}, a phenomenon referred to as the dissipative anomaly.
A similar phenomenon is observed in the kinetics of wave turbulence \cite{CON2009}. 
There is no physical contradiction in the fact that mass conservation is
broken in the Smoluchowski equation in the gelling regime. It simply means  that
 for times larger than the gelation time, the underlying conservative coagulation 
dynamics must be modified for the largest clusters. 

\begin{figure}
\begin{center}
\includegraphics[width=7.0cm]{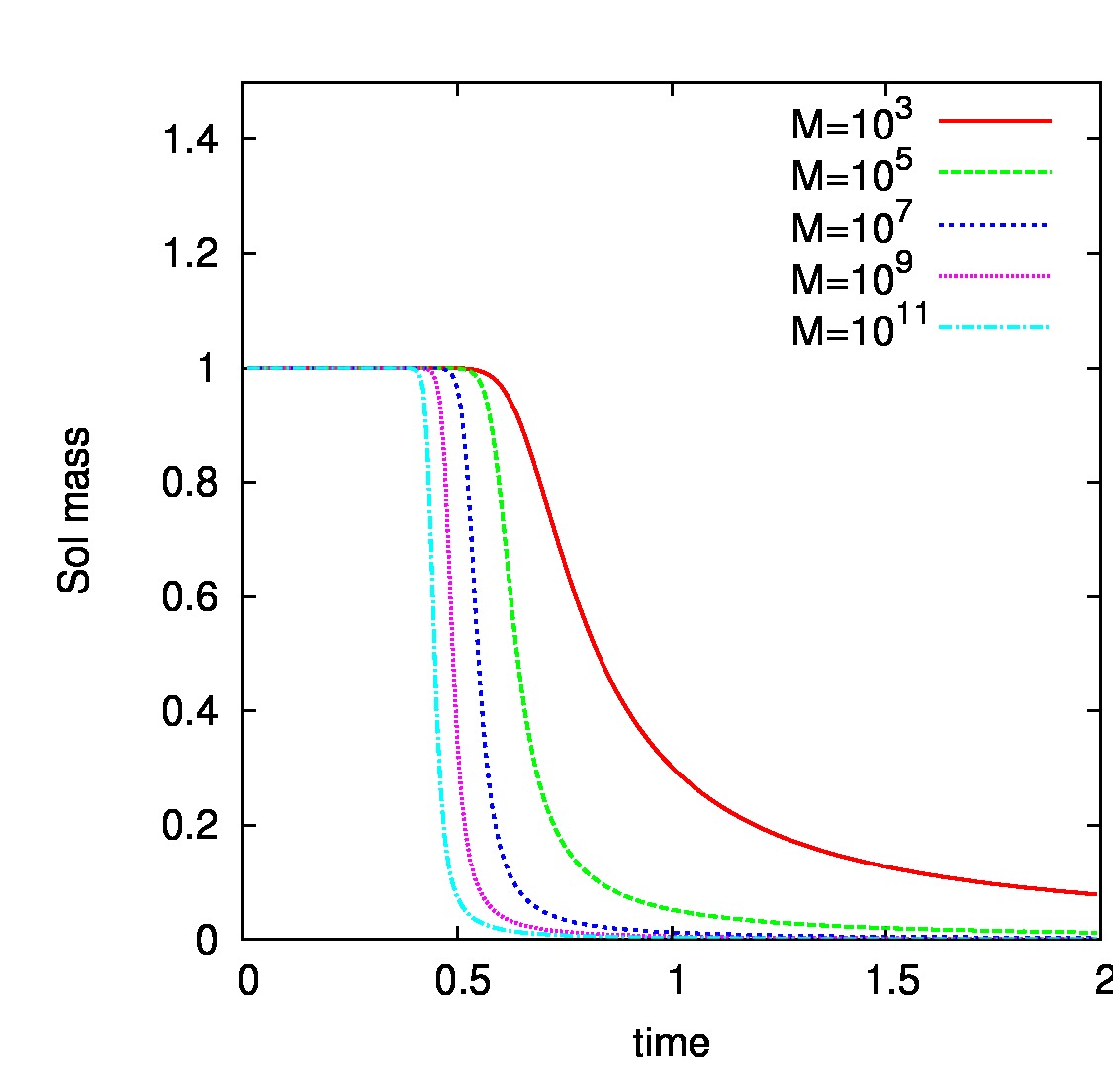}\\
\caption{\label{fig-IG}(Color online)  Total mass density, $\M_1(t)$, for the generalised sum 
kernel, $K(m_1,m_2)=m_1^\frac{3}{2} + m_2^\frac{3}{2}$.}
\end{center}
\end{figure}

Let us now consider what happens when an instantaneously gelling kernel is 
inserted into the regularised Smoluchowski equation. An archetypal instantaneously
gelling kernel, which we shall use extensively in what follows,  is the 
generalised sum kernel:
\begin{equation}
\label{eq-genSumKernel}
K_\epsilon(m_1,m_2) = m_1^{1+\epsilon} + m_2^{1+\epsilon}.
\end{equation}
For this kernel $\lambda = \nu = 1+\epsilon$ and $\mu=0$. According to the 
classification of Van Dongen and Ernst, it is nongelling for $\epsilon<0$ and 
instantaneously gelling for $\epsilon>0$. In the marginal case, $\epsilon=0$, it is the
simple sum kernel which is exactly solvable, at least in the absence of a
source of mononers \cite{LEY2003}, and turns out to
be non-gelling. Fig.~\ref{fig-IG} shows the sol mass in the regularised system
with the generalised sum kernel with $\epsilon=\frac{1}{2}$ for a sequence of
increasing values of the regularisation mass, $M$. In the 
presence of the cut-off, the regularized gelation time, $t^*_M>0$, 
is clearly identifiable. 
This regularised gelation time, although finite, {\em decreases} as  $M_{\mathrm max}$ {\em increases}. 
Extrapolating the behaviour seen in Fig.~\ref{fig-IG}, it is plausible that
the gelation time vanishes as $M \to \infty$ consistent with the
expectation that the unregularised system exhibits complete instantaneous
gelation.  Instantaneous gelation has not, to the best of our knowledge, been
successfully numerically demonstrated in the literature previously. In the most
extensive numerical study of the Smoluchowski equation to date, that of  Lee 
\cite{LEE2001}, the numerical difficulties posed by kernels like 
Eq.~(\ref{eq-genSumKernel}) were explored and it was concluded that there are
no self-consistent solutions of Eq.~(\ref{eq-SKE}) for such kernels. Our 
results demonstrate that the non-conservative regularisation, 
Eq.~(\ref{eq-regSKE}), provides one way around these difficulties, at least
numerically, although we expect that the regularisation could also be useful for
rigorous mathematical studies.

From a physical point of view, the most important observation about the 
results presented in Fig.~\ref{fig-IG}
is that the regularised gelation time decreases extremely slowly as the
cut-off is increased. $t^*_M$ decreases by a factor of less than
2 as $M$ is increased by 8 orders of magnitude.  This very weak 
dependence means that, in practice, kernels which would exhibit instantaneous
gelation, even complete instantateous gelation, for $M=\infty$,
can still have smooth, physically reasonable behaviour for finite $M$.
The regularised gelation time, $t^*_M$, depends sufficiently weakly on the actual 
value of $M$ that such regularised systems may still be  useful in modelling. On 
the basis of our numerics, we
conjecture that as $M\to \infty$, the regularised gelation time decreases as 
$t^*_M \sim (\log M)^{-\alpha}$ for some $\alpha>0$. 
This would complement heuristic
arguments put forward by Ben-Naim and Krapivsky \cite{BNK2003} for gelation in
finite systems of particles undergoing exchange-driven growth for which it is
argued that the gelation time decreases as a power of the logarithm of the
initial number of particles when the aggregation rate increases sufficiently
quickly as a function of the mass of the larger cluster and thus becomes 
instantaneous in the limit of an infinite number of initial particles. 
An important question, which is left open, is to develop an analytic approach
allowing the determination of the functional dependence of $t^*_{M_\mathrm{max}}$ 
on $M_\mathrm{max}$ and the value of $\alpha$ if the dependence conjectured
above in indeed correct. We feel it is unlikely that the numerics can be extended to 
sufficiently large values of $M_\mathrm{max}$ to determine this dependence
unambiguously although on the basis of the numerics we have available and
taking into account the analytic work on the corresponding problem with a 
source of monomers reported below, we would not be surprised if $\alpha=1/2$.

We devote much of the remainder of the paper to studing what happens when a source of
monomers is introduced into a system with an instantaneously gelling kernel.
Such a situation has been partially analysed by Kontorovich \cite{KON01} 
in the context of gravitational clustering and by Horvai et al. \cite{HNS2008} in
the context of differential sedimentation. Both studies concluded that the
system should reach a stationary state at large times in which injection 
of monomers is balanced by the aggregation of smaller clusters into
larger ones (a mass ``cascade''). In both cases, the cascade is non-local
in the sense that the transfer of mass from small clusters to large is
dominated by the interaction of very large and very small clusters. For a
detailed discussion of the criteria for locality of mass transfer in 
cascade solutions of the Smoluchowski equation see \cite{CRZ2004}.

\section{The addition model: a simplified description of runaway absorption of small clusters by large ones}
\label{sec-additionModel}
\begin{figure}
\begin{center}
\includegraphics[width=7.0cm]{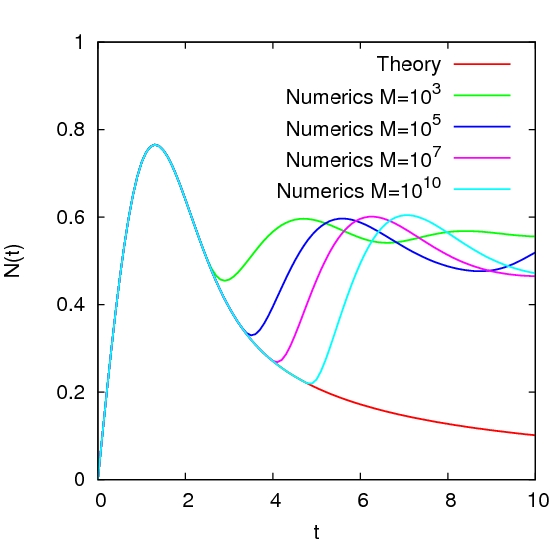}\\
\caption{\label{fig-N}(Color online) Total particle density, $\M_0(t)=\int_0^\infty N_m(t)\,dm$  for $K(m_1,m_2)=m_1+m_2$ and source.}
\end{center}
\end{figure}

Instantaneous gelation is driven by the runaway absorption of small 
clusters by large ones.  This fact is most easily seen from the analytically 
tractable (but non-gelling) marginal kernel, $m_1+m_2$, with source of monomers.
For $\epsilon=0$, the kernel in Eq.(\ref{eq-genSumKernel}) is the standard sum kernel.
In this case, integrating Eq.(\ref{eq-SKE}) with respect to $m$ 
(we have set $m_0=1$) allows us to obtain the following equation for the total number of particles in the system:
\begin{equation}
\dd{N}{t} = - 2 \M_1(t) N(t) + J.
\end{equation}
There is no gelation transition at finite time in this case so
$\M_1(t) = J t$ and we get a closed equation for $N(t)$:
\begin{equation}
\dd{N}{t} = - 2 J t N(t) + J.
\end{equation}
The solution is
\begin{equation}
\label{eq-NSumKernel}
N(t) = \frac{1}{2} \sqrt{J \pi}\ {\mathrm{Erfi}}(\sqrt{J}\, t)\ {\mathrm e}^{-g J t^2},
\end{equation}
where $\mathrm{Erfi}(x) = -\frac{2 i}{\sqrt{\pi}} \int_0^{ix} \mathrm{e}^{-y^2} dy$ is the ``imaginary error function''.
This formula shows an interesting feature which is surprising at first sight: 
while the total number of particles in the system initially increases as we add 
particles, it subsequently reaches a maximum and starts to decrease as shown
in Fig. \ref{fig-N}. It tends to 
zero ($\asymp 1/t$) as time gets large. Meanwhile the total mass in the system 
increases linearly. This tells us that the long time behaviour of the system is 
dominated by an ever decreasing number of increasingly massive particles which 
immediately eat all the monomers injected into the system.  Thus we already see 
the essential feature of non-local interactions at the level of the sum kernel. 
Fig. \ref{fig-N} demonstrates this behaviour. The numerical solution follows the 
analytic prediction, Eq.~(\ref{eq-NSumKernel}), for longer and longer times as the 
mass cut-off $M$ is increased.
This case is marginal in the sense that there is no finite time gelation even 
though all the mass gets concentrated in larger and larger clusters. If 
$\epsilon>0$, and the exponent $\nu>1$, then big clusters are so ``hungry" that 
they consume all smaller clusters at a rate which diverges with the cut-off, 
$M_\mathrm{max}$. 

The dynamics of a coagulation system dominated by the absorption of small clusters
by large ones is modelled in extremis by the so-called ``addition model" 
(Brilliantov \& Krapivsky \cite{BK1991}) in which clusters are only permitted to
grow by reacting with monomers. Many of the interesting phenomena which we observe in
the full coagulation problem have qualitative analogues for the addition model
so we shall devote some time to studying it. The addition model corresponds to the kernel
\begin{equation}
\label{eq-additionModelKernel}
K(m_1, m_2) =  \frac{1}{2}\left(\kappa(m_1)\, \delta(m_2-1) + \kappa(m_2)\,\delta(m_1-1)\right),
\end{equation}
where the function $\kappa(m)$ is typically taken to be a homogeneous function of
degree $\nu$. This considerably simplifies the coagulation dynamics. 
Although the addition model can be analyzed for a fairly general set of gelling reaction rates
$\{\kappa(m)\}_{m\geq 1}$, we will use the following example for illustrative purposes:
\bea
\label{eq-modelAMKernel}
\kappa(m)=\left\{
\begin{array}{ll}
\gamma& m=1\\ 
m^{1+\epsilon}, &m=1,2,\ldots, M,~ \epsilon >0.\label{eq:cc}
\end{array}\right.
\eea
Here we have changed notation slightly and written $\nu=1+\epsilon$ so that the instantanenous gelation
regime corresponds to $\epsilon>0$. The reason for introducing a separate parameter, $\gamma$, to describe
the monomer reaction rate will become clear in Sec.~\ref{sec-oscillationsInAM} where we show that by choosing 
$\gamma$ appropriately, it is possible to make the connection between the addition model and the Smoluchowski 
equation more quantitative.  We shall also impose the initial condition
\bea
N_m(t)=N_0\delta_{m,1}, \label{eq:ic}
\eea
excluding the possibility of zero initial momomer concentration (this is to regularize a change of time to be 
defined below).

Insertion of Eqs.~(\ref{eq-additionModelKernel}) and (\ref{eq-modelAMKernel}) into Eq.~(\ref{eq-SKE}) yields the
discrete set of equations
\bea
\frac{dN_1}{dt}(t)&=&-2\,\gamma\,N_1^2-\sum_{m=2}^M \kappa(m)N_1 N_m+J,\label{eq:m1}\\
\nonumber \frac{dN_m}{dt}(t)&=&\kappa(m-1)N_1N_{m-1}-\kappa(m)N_1 N_m\\
& &~2\leq m\leq M,\label{eq:mm}
\eea
where explicit dependence of the $N_m$ on $t$ has been suppressed for brevity and a cut-off, 
$M$, has been introduced.
Instantaneous gelation in the addition model without a cut-off for $\nu>1$ in the absence of a 
source was conjectured in \cite{BK1991} and subsequently proven by Laurencot 
\cite{LAU1999} providing support for our intuition that the process of
aggregation of monomers by large particles is a runaway process in the
full aggregation equation when $\nu>1$. Before continuing, we would like to 
emphasise again that we only expect the addition model to be similar to the Smoluchowski 
dynamics in the instantaneously gelling regime when the dynamics is dominated by the aggregation of
large clusters and monomers. It is known that the addition model
has very different dynamics to the Smoluchowski equation in the non-gelling regime,  $\nu<1$. In 
particular it does not exhibit scaling in the absence of a source \cite{BK1991}.

Eqs. (\ref{eq:m1}, \ref{eq:mm}) simplify after a non-linear change of physical time to ``monomer'' time $\tau$ defined
as follows:
\bea
d\tau(t)&=&N_1(t)dt, \label{eq:physt}\\
\tau (0)&=&0.
\eea
Once the solution parameterized by the monomer time has been found, the inverse map to the physical time is given by
\bea
t(\tau)=\int_{0}^{\tau}\frac{d\tau'}{N_1(\tau')}
\eea
This integral is proper at $\tau=0$ due to the initial condition, Eq.~(\ref{eq:ic}).
In terms of the monomer time, Eqs. (\ref{eq:m1}, \ref{eq:mm}) take the following form:
\bea
N_1^\prime(\tau)&=&-2\,\gamma\,N_1(\tau)-\sum_{m=2}^M \kappa(m)N_m+\frac{J}{N_1},\label{eq:m1mt}\\
\nonumber N_m^\prime(\tau)&=&\kappa(m-1)N_{m-1}-\kappa(m) N_m,\\
& &~2\leq m\leq M,\label{eq:mmmt}
\eea
where $^\prime$ denotes the monomer time derivative $\frac{d}{d\tau}$.
Note that Eq.~(\ref{eq:mmmt}) is an inhomogeneous linear system of ODE's with the inhomogeneity being a nonlinear function of
the monomer concentration $N_1(\tau)$. It can be solved with respect to the concentrations of polymers using
the Laplace transform in the monomer time.
The solution can be written in the form
\bea
\nonumber N_m(\tau)&=&\int_{0}^{\infty} d\tau' K_{m}(\tau-\tau') N_{1}(\tau'),\\
& & m=1,2,\ldots, M,\label{eq:poly}
\eea
where
\bea
K_m(\tau)=\gamma\,\kappa(m)^{-1}\int_{\Gamma} \frac{d\omega}{2\pi} e^{\omega \tau}\prod_{k=2}^m \frac{1}{1+\frac{\omega}{\kappa(k)}},
\label{eq:ker}
\eea
the integration contour $\Gamma=(\sigma-i\infty,\sigma+i\infty)$, and $\sigma$  is a real constant:
$\sigma>-\min_{m\geq 2} [ \kappa(m) ]$.
It is easy to see from Eq.~(\ref{eq:ker}) that
$$
K(\tau)=0, ~\tau<0,
$$
which is an expression of causality in the addition model.
Substituting (\ref{eq:poly}) in (\ref{eq:m1mt}) we arrive at a nonlinear integral
equation for the monomer concentration:
\bea
\nonumber N_1'(\tau)&=&-2\,\gamma\,N_1(\tau)- \kappa(1)\,\int_0^{\tau} d\tau' Q_M(\tau-\tau')N_1(\tau')\\
&&+\frac{J}{N_1(\tau)},\label{eq:me}
\eea
where
\bea
Q_M(\tau)=\sum_{m=2}^M \int_{\Gamma} \frac{d\omega}{2\pi} e^{\omega \tau} \prod_{k=2}^m \frac{1}{1+\frac{\omega}{\kappa(k)}}
\label{eq:kers}
\eea
It is clear from Eq.~(\ref{eq:me}) that $N_1(\tau)$ stays positive at all times. The transformation to monomer
time is therefore well defined.
To find the steady state solution of Eq.~(\ref{eq:me}) we note that all the poles in the integrand of Eq.~(\ref{eq:kers})
have negative real parts. Thus,
\[
\lim_{\tau\rightarrow \infty} \int_{\tau} d\tau' Q_M(\tau-\tau')N_{1}(\tau')=N_{1}(\infty)\int_0^\infty d\tau Q_M(\tau).
\]
Substituting this answer into Eq.~(\ref{eq:me}) and setting $N'(\infty)=0$ we find that
\bea
\nonumber N_1(\infty)&=&\sqrt{\frac{J}{\gamma\left(2+\int_0^\infty Q_M(\tau)\,d\tau\right)}}\\
&=&\sqrt{\frac{J}{\gamma\,\left(M+1\right)}},
\eea
where the last equality results from direct integration of the kernel,  Eq.~(\ref{eq:kers}), over time
using Cauchy's theorem.
For the specific case of the kernel, Eq.~(\ref{eq:cc}), the corresponding steady state answer for the polymer 
concentrations is
\bea
\nonumber N_m(\infty) &=& \gamma\,N_1(\infty)\, \kappa(m)^{-1}\\
&=& \sqrt{\gamma}\,\sqrt{\frac{J}{M+1}}\,m^{-1-\epsilon}.
\label{eq-ssAM}
\eea
We acknowledge that it would be easier to find the steady state directly from Eqs.~(\ref{eq:m1}, \ref{eq:mm}), but 
Eq.~(\ref{eq:me}) will be better suited to study the time evolution in what follows.

\section{The full coagulation problem with $\nu>1$ in the presence of a source of monomers}
In the presence of a source of monomers, Eq.~(\ref{eq-SKE}) has a formal 
stationary solution which scales for large masses as 
$N_m \sim m^{-\frac{\lambda+3}{2}}$ which describes a cascade of mass from small
masses to large \cite{CRZ2004}. This solution is only valid if the collision integral 
is convergent. For the case of the kernel Eq.~(\ref{eq-genSumKernel}), this
convergence criterion fails for $\epsilon>0$ \cite{CRZ2004} presumably reflecting
the tendency for the system to gel instantaneously in this regime.
When a cut-off is introduced, a stationary state may be reached if a source
of monomers is present although this stationary state must involve the cut-off 
rather than being of the cascade type. In \cite{HNS2008} such a stationary state was
found for the case of differential sedimentation ($\nu=\frac{4}{3}$) which
scaled for large masses as $N_m \sim m^{-\frac{4}{3}}$.

\begin{figure}
\begin{center}
\includegraphics[width=7.0cm]{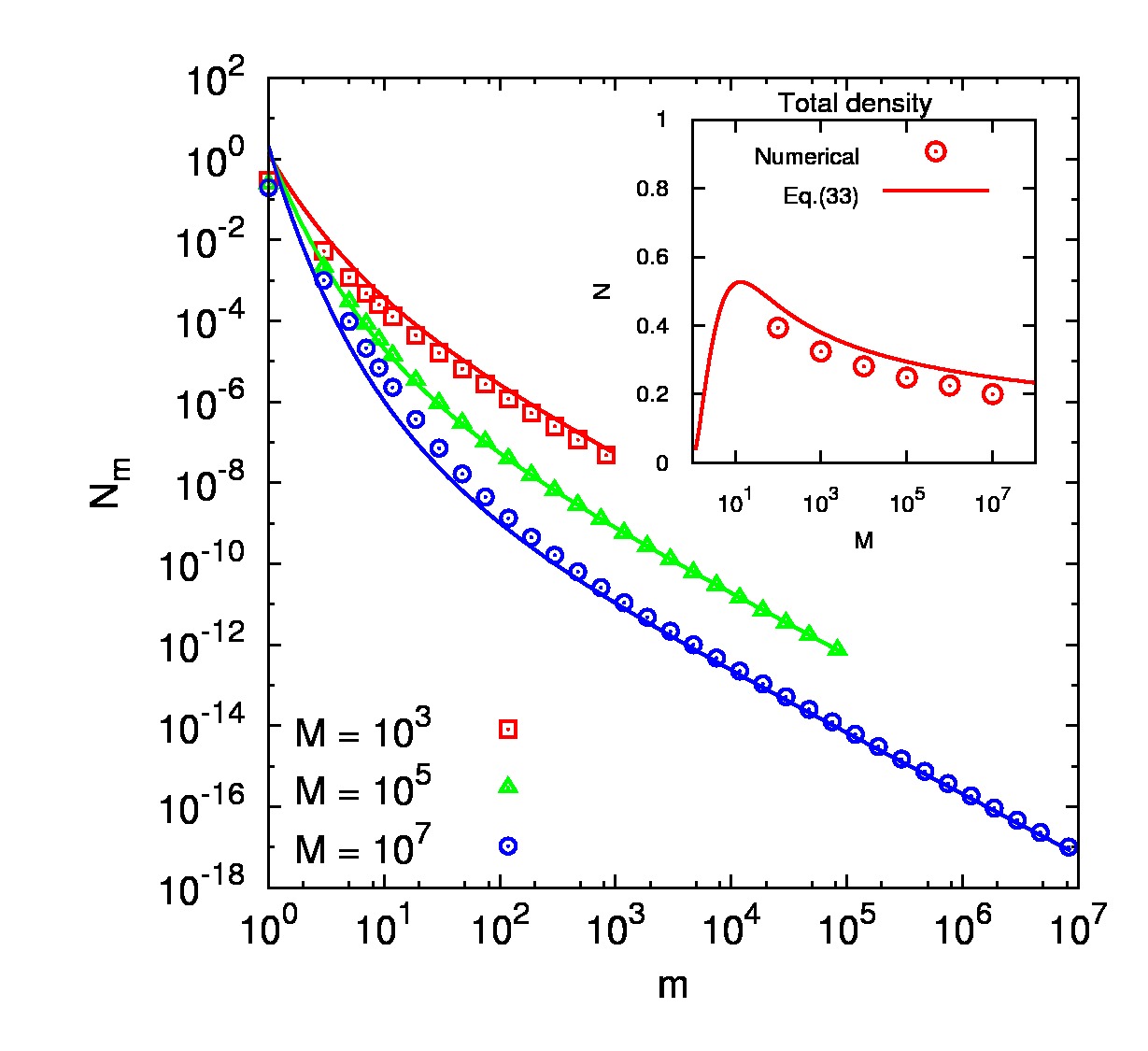}\\
\vspace{-0.25cm}
\caption{\label{fig-stationaryState}(Color online) Non-local stationary state (theory vs numerics) for $\nu=3/2$.}
\end{center}
\end{figure}

In this section we perform a systematic analysis of the stationary state
for the model kernel
\begin{equation}
\label{eq-maxKernel}
K(m_1,m_2)=\mathrm{max}(m_1, m_2)^\nu,
\end{equation}
which captures the essential features of instantaneously gelling systems
while permitting some convenient simplifications of the equations. If we assume 
that the system reaches a stationary state and define
\begin{equation}
\Gamma_M=\sum_{m=1}^{M}m^{\nu}N_{m},
\end{equation}
then some manipulations of the discrete analogue of Eq.~(\ref{eq-SKE}) show that 
the stationary state can be expressed via a recursion relation:
\begin{equation}
N_{m}=\frac{\frac{1}{2}\sum_{p=1}^{m-1}\max(p,m-p)^{\nu}N_{p}N_{m-p}}{\Gamma_M+\sum_{p=1}^{m-1}\left(m^{\nu}-p^{\nu}\right)N_{p}}.
\end{equation}
which can be iterated to find $N_m$ once one observes that the monomer concentration
is fixed in the stationary state as:
\begin{equation}
N_1 = \frac{J}{\Gamma_M}
\end{equation}
For each value of $M$, this iteration procedure can, in principle, be used to
self-consistently determine $\Gamma_M$. In this paper, we adopt a different 
approach which is based on the assumption that the mass transfer is nonlocal
in mass space. Following the approach of \cite{HNS2008}, Eq.~(\ref{eq-SKE})
can be approximated by:
\begin{equation}
\label{eq-nonlocalSKE}
\pd{N_m}{t} = -\M_1 \pd{}{m}\left(m^\nu N_m\right) - \M_\nu\,N_m 
\end{equation}
where the moments $\M_1$ and $M_\nu$ are now to be computed with the cut-off retained:
\begin{equation}
\label{eq-moment}
\M_\alpha = \int_1^M m^\alpha\,N_m\,d m.
\end{equation}
The source and sink terms have been omitted for now. In obtaining 
Eq.~(\ref{eq-nonlocalSKE}) it has been assumed that the integrals $\M_1$ and
$\M_\nu$ are dominated by their lower and upper limits of integration
respectively. The consistency of this assumption will be determined a-posteriori.
Since $\M_1$ and $\M_\nu$ do not depend on $m$, Eq.~(\ref{eq-nonlocalSKE})  has the stationary solution:
\begin{equation}
\label{eq-naiveSolution}
N_m = C\,\exp\left[ \beta \frac{m^{1-\nu}}{\nu-1}\right]\,m^{-\nu},
\end{equation}
where $C$ is an arbitrary constant and we have, for convenience, introduced the 
parameter,
\begin{equation}
\label{eq-beta}
\beta = \frac{\M_\nu}{\M_1}.
\end{equation}
Although we do not know the value of $\beta$ a-priori, it can now be determined
self-consistently due to the fact that, when $N_m$ is given by 
Eq.(\ref{eq-naiveSolution}), the integrals $\M_1$ and $\M_\nu$ can be
expressed explicitly in terms of incomplete gamma functions:
\begin{widetext}
\begin{eqnarray}
\label{eq-M1} \M_1(\beta) &=& \frac{C}{\nu-1} \left(\frac{1-\nu}{\beta}\right)^\frac{\nu-2}{\nu-1}\left[
 \Gamma\left(\frac{\nu-2}{\nu-1}, -\frac{\beta\,M^{1-\nu}}{\nu-1} \right)
-\Gamma\left(\frac{\nu-2}{\nu-1}, -\frac{\beta}{\nu-1} \right) \right]\\
\label{eq-Mnu} \M_\nu(\beta) &=& \frac{C}{\nu-1} \left(\frac{1-\nu}{\beta}\right)^\frac{1}{1-\nu}\left[ 
 \Gamma\left(\frac{1}{1-\nu}, -\frac{\beta\,M^{1-\nu}}{\nu-1} \right)
-\Gamma\left(\frac{1}{1-\nu}, -\frac{\beta}{\nu-1} \right) \right].
\end{eqnarray}
After some algebra, Eq.~(\ref{eq-beta}) reduces to the following consistency
condition for the value of $\beta$:
\begin{equation}
\label{eq-consistency}
\Gamma\left(\frac{\nu-2}{\nu-1}, -\frac{\beta\,M^{1-\nu}}{\nu-1} \right)
-\Gamma\left(\frac{\nu-2}{\nu-1}, -\frac{\beta}{\nu-1} \right)
=-\frac{1}{\nu-1}\,\Gamma\left(\frac{1}{1-\nu}, -\frac{\beta\,M^{1-\nu}}{\nu-1} \right)
+\frac{1}{\nu-1}\,\Gamma\left(\frac{1}{1-\nu}, -\frac{\beta}{\nu-1} \right)
\end{equation}
\end{widetext}
Given $\nu>1$, this can be solved numerically for any value of $M$. Such
a numerical investigation indicates that $\beta$ is a slowly increasing function
of $M$.  Furthermore, since we expect the integrals $\M_1$ and $\M_\nu$ to be
dominated by their respective lower and upper limits of integration as $M\to \infty$, we
would expect
Eq.~(\ref{eq-consistency}) to satisfy the following asymptotic balance for large 
$M$:
\begin{equation}
\label{eq-balance}
-\Gamma\left(\frac{\nu-2}{\nu-1}, -\frac{\beta}{\nu-1} \right) \sim -\frac{1}{\nu-1}\,\Gamma\left(\frac{1}{1-\nu}, -\frac{\beta\,M^{1-\nu}}{\nu-1} \right).
\end{equation}
Since numerics indicate that $\beta$ increases (but only slowly) as $M$ grows and 
$\nu>1$, the argument of the left-hand gamma function, $\frac{\beta}{\nu-1}$,
should increase in the limit of interest. Likewise, the argument of the right-hand 
gamma function, $\frac{\beta\,M^{1-\nu}}{\nu-1}$, should decrease in the limit of
interest. The relevant leading order asymptotics are then:
\begin{eqnarray}
\label{eq-gammaAsymptotics2} \Gamma(a, z) &\sim z^{a-1}\, e^{-z} &\mbox{as $z\to \infty$}\\
\label{eq-gammaAsymptotics1} \Gamma(a, z) &\sim -\frac{1}{a}\,z^a &\mbox{as $z\to 0$}.
\end{eqnarray}
Substituting these into Eq.~(\ref{eq-balance}), one finds that the leading order
terms balance as $M \to \infty$ provided
\begin{equation}
\label{eq-betaValue}
\beta \sim \log(M^{\nu-1}).
\end{equation}
Whilst the index $\nu-1$ could be absorbed within the implied front factor in 
$\beta$, we keep it in the following expressions since it captures how the 
dependence on the cut-off, $M$, goes away when $\nu$ approaches unity.
Knowing $\beta$, Eqs.~(\ref{eq-gammaAsymptotics1}) and (\ref{eq-gammaAsymptotics2})
can now be used to detemine the asymptotic behaviour of $\M_1$ and $\M_\nu$
from Eqs.~(\ref{eq-M1}) and (\ref{eq-Mnu}). After some work one finds:
\begin{eqnarray}
\label{eq-M1-2} &\M_1 \sim \frac{C M}{\log(M^{\nu-1})} &\mbox{as $M\to \infty$}\\
\label{eq-Mnu-2} &\M_\nu \sim C\, M &\mbox{as $M\to \infty$}.
\end{eqnarray}
It now remains to determine the constant $C$ in Eq.~(\ref{eq-naiveSolution})
in order to close the argument and allow us to check for consistency.
This can be done using global mass balance. Multiplying 
Eq.~(\ref{eq-regSKE}) (with the kernel Eq.~(\ref{eq-maxKernel})) by $m$ and 
integrating from 1 to $M$ yields the global mass balance condition:
\begin{equation}
J = \int_1^{M} d m \ m\,N_m \int_{M-m}^M d m_1\ m_1^\nu N_{m_1}.
\end{equation}
Our assumption of nonlocality allows to replace the inner integral
by $\M_\nu$ in the limit of large $M$ since it is dominated by its upper 
limit. Consequently the global mass balance condition in the limit of large $M$ is
\begin{equation}
J = \M_1\, \M_\nu = \frac{C^2 M^2}{\log(M^{\nu-1})},
\end{equation}
where we have used Eqs.~(\ref{eq-M1-2}) and (\ref{eq-Mnu-2}). This gives
\begin{equation}
\label{eq-amplitude}
C = \frac{\sqrt{J \log(M^{\nu-1})}}{M}.
\end{equation}
Putting this all together with Eq.~(\ref{eq-naiveSolution}) the asymptotic solution of Eq.~(\ref{eq-nonlocalSKE}) is
\begin{equation}
\label{eq-nonlocalSolution}
N_m = \frac{\sqrt{J\,\log(M^{\nu-1})}}{M}\, M^{m^{1-\nu}} \, m^{-\nu}.
\end{equation}
Everything becomes self-consistent: $\mathcal{M}_1$ becomes independent of $M$ and
 $\mathcal{M}_\nu$ diverges as $\sqrt{\log(M^{\nu-1})}$ when $M$ gets
large thereby justifying the use of Eq.~(\ref{eq-nonlocalSKE}).
This theory does a good job of capturing the general features of the stationary
state: stretched exponential decay at small cluster sizes followed by a cross-over to
a power law decay with exponent $\nu$ for large cluster sizes but with an amplitude which
decreases with the cut-off, $M$. Fig.~\ref{fig-stationaryState} compares
Eq.~(\ref{eq-nonlocalSolution}) with the stationary state obtained from
numerical simulations of Eq.~(\ref{eq-regSKE}) with $\nu=3/2$ for a range of
values of $M$. The agreement is excellent given that there are no
adjustable parameters in Eq.~(\ref{eq-nonlocalSolution}). Note that the large mass
behaviour, $N_m \sim m^{-\nu}$, is in agreement with the stationary solution of the
addition model, Eq.~(\ref{eq-ssAM}), obtained in Sec.~\ref{sec-additionModel}.

Finally, we can now quantify the rate at which the system becomes singular
as the cut-off, $M$, is increased. Using Eq.~(\ref{eq-nonlocalSolution}) to
calculate the total particle density in the stationary state we get
\begin{equation}
\label{eq-densityDecay}
N = \frac{\sqrt{J} \left(M -M^{M^{1-\nu}} \right)}{M\,\sqrt{\log(M^{\nu-1})}} \sim\sqrt{\frac{J}{\log(M^{\nu-1})}} \  \ \mbox{as $M \to \infty$}.
\end{equation}
The density of particles in the stationary state thus vanishes as the cut-off is 
removed which is the signature of instantaneous gelation. The approach to zero
is logarithmically slow however. The asymptotic estimate,
Eq.~(\ref{eq-densityDecay}), is compared against the total density measured
from a sequence of numerical simulations with $\nu=3/2$ and different values of 
the cut-off in the inset of Fig.~\ref{fig-stationaryState}. The theory
again produces convincing agreement with numerics without any adjustable 
parameters.

\section{Approach to the Stationary State}
\label{sec-oscillations}
\begin{figure}
\begin{center}
\includegraphics[width=7.0cm]{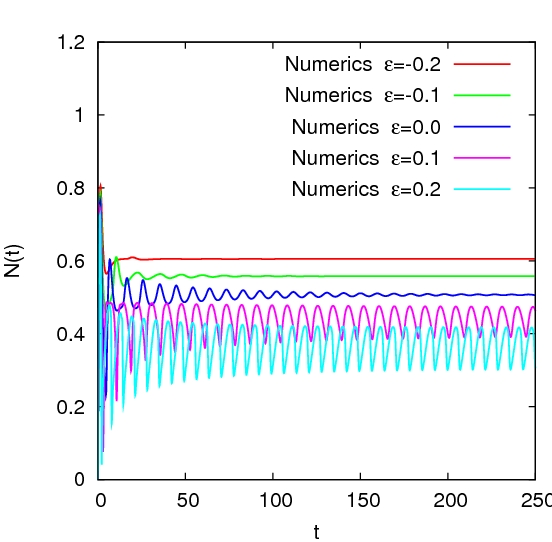}\\
\caption{\label{fig-oscillations}(Color online) Total density vs time for $K(m_1,m_2)=m_1^{1+\epsilon}+m_2^{1+\epsilon}$.}
\end{center}
\end{figure}

\begin{figure}
\begin{center}
\includegraphics[width=7.0cm]{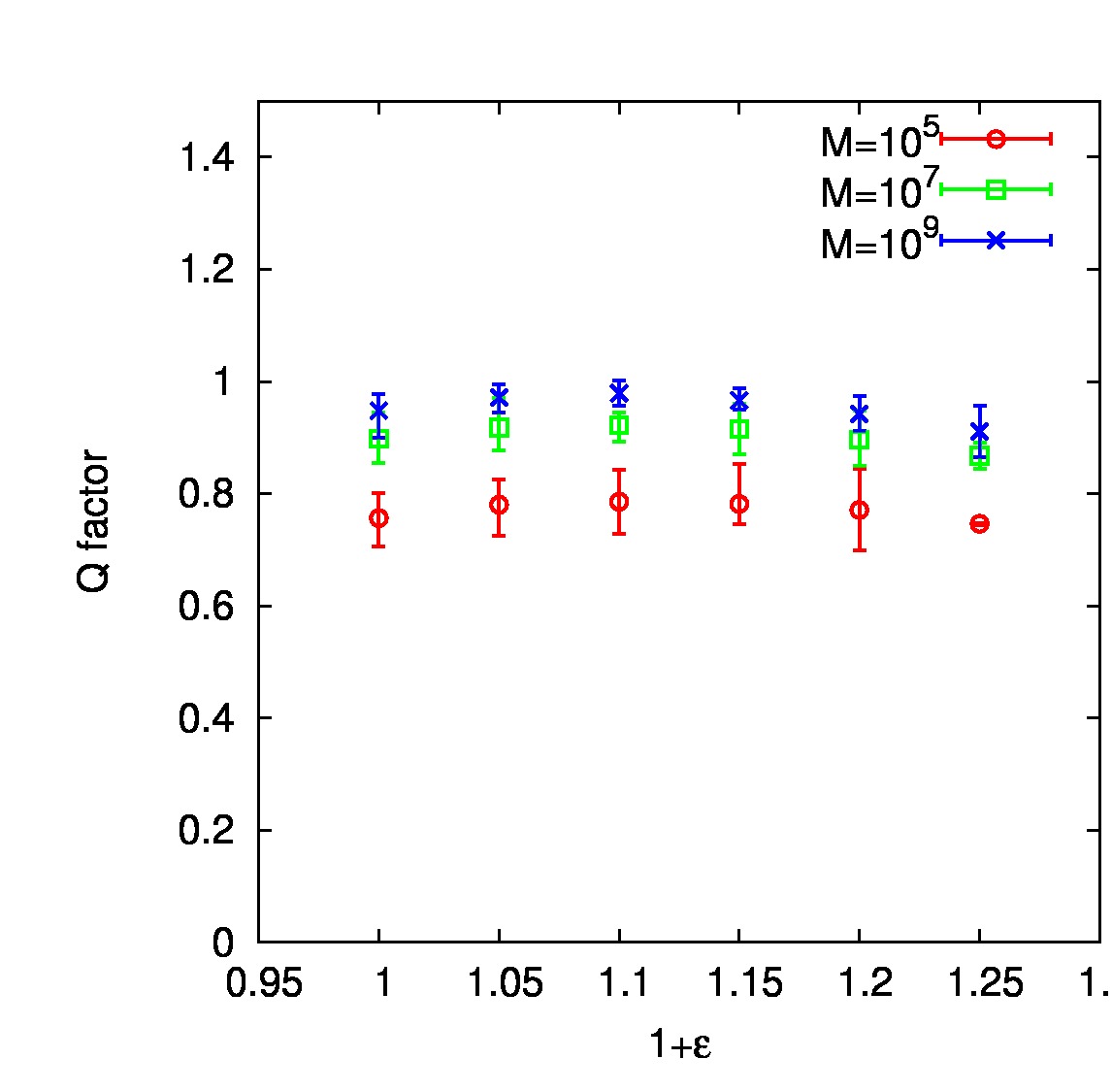}\\
\caption{\label{fig-Qfactor}(Color online) ``Q-factor" of the oscillatory transient for a range of values of $\nu=1+\epsilon$ and a range of cut-offs, $M$.}
\end{center}
\end{figure}

Numerical simulations indicate that the approach to the stationary state
is interesting. Fig.~\ref{fig-oscillations}
shows a plot of the total density in the system as a function of time for the
generalised sum kernel, Eq.~(\ref{eq-genSumKernel}), for a range of values of
 $\epsilon$ with cut-off $M=10^9$. In the nonlocal regime, 
$\epsilon >0$, the approach to the stationary state is characterised by long-lived
transient oscillations of the density in the system. This is surprising since
there is nothing of an oscillatory or cyclic character in the underlying 
coagulation dynamics. {We note, however, that long-lived oscillations
of the total droplet density have been observed in experiments on the phase
separation of binary mixtures with nucleation and coagulation of droplets 
\cite{VAV2007,BV2010} which may be related to the effect noted here. From Fig.~\ref{fig-oscillations}, it is not obvious that
these oscillations are decaying at all for $\epsilon>0$. From our numerical
investigations we believe that these oscillations are indeed transient although
the rate of decay becomes very slow as the cut-off is increased. Although the
oscillations are quite nonlinear in character, we defined a quantity analogous
to the Q-factor of a linear oscillator by measuring the ratio of subsequent
maxima of the signal and defining the Q-factor to be the estimated numerical
limit of this ratio as $t \to \infty$ (for all of our numerical experiments with
$\epsilon >0$, we found that this ratio does indeed seem to converge to a
finite value). Thus a Q-factor of 1 would correspond to persistent oscillations.

A summary of these numerical experiments is presented in Fig.~\ref{fig-Qfactor}
which shows numerically estimated Q-factors for the oscillations observed for
a range of values of $\epsilon$ and $M$. The most evident trend 
from the data is the fact that the oscillations come closer and closer to a
Q-factor of 1 as $M$ is increased. These measurements indicate
that the oscillations are decaying in all cases but only very slowly when
the cut-off becomes large. The behaviour as a function of $\epsilon$ is less
clear with some possible evidence for a weak maximum.

We offer the following heuristic explanation for how oscillations can be generated
in this system. For $\epsilon>0$, the case for which instantaneous gelation should 
occur in the absence of a cut-off, the indication of Fig. \ref{fig-IG} is that the 
particle density drops close to zero in finite time (but this time is not zero). 
When monomers are injected into the system, until the cut-off is felt, the mass in 
the system just increases linearly. However, as soon as large particles are 
generated, their absorption of the monomers is a runaway process which 
very rapidly converts the monomers which have accumulated in the system into
large particles which are immediately removed by the cut-off. This then
resets the system close to its initial state in which there are almost no particles
in the system. The dynamics then repeats. 

\section{Approach to the stationary state in the addition model}
\label{sec-oscillationsInAM}
We do not, as yet, have an analytic characterisation of the oscillations described in the previous
section. Instead, we conclude this article with an analysis of the analogous
problem for the simpler case of the addition model discussed in Sec.~\ref{sec-additionModel}
for which the oscillatory approach to the stationary state, Eq.~(\ref{eq-ssAM}),
can be derived.

The addition model aims to model the most important contribution to the dynamics of the full aggregation
model with non-local kernels by concentrating on interactions between ``light'' and ``heavy'' particles only.
At the level of modeling, there is no reason to expect that the parameters of the addition model are the
same as the analogous parameters in the full model. Before proceeding with calculations, we first present
a semi-heuristic argument for the most appropriate choice of parameters.
Let us first clarify what is meant by ``light'' and ``heavy'' in terms of the steady state of the full model,
Eq.~(\ref{eq-nonlocalSolution}). Heavy particles have masses in the polynomial tail of 
Eq.~(\ref{eq-nonlocalSolution}), $m\gg \log(M)^{\nu-1}$. Light particles
belong to the stretched exponential region of Eq.~(\ref{eq-nonlocalSolution}), 
$m\ll \log(M)^{\nu-1}$. Therefore, neglecting logarithmic corrections, {\em we can treat all light particles as 
effective monomers.} Hence, the parameters $J$ and $\kappa_m$
for $m>1$ in the addition model are the same as in the full model, but the parameter $\gamma=\kappa(1)$
should be chosen as an effective interaction rate of all light particles in the full model with themselves. The 
easiest way to choose this rate is to require that the steady state of the addition model gives the correct 
description of the mass distribution of the full model at large masses.
Comparing Eq.~(\ref{eq-ssAM}) with the large-$m$ behaviour of Eq.~(\ref{eq-nonlocalSolution}) we see that we
should choose:
\bea
\gamma\sim \frac{1}{M}.
\eea

For large times let us look for a solution of Eq.~(\ref{eq:me}) in the form
\bea
N_1(\tau)=\sqrt{\frac{J}{\gamma\left(M+1\right)}}(1+Ae^{\zeta\tau}),
\label{eq:anz}
\eea
where $\mathrm{Re}\,\zeta<0$. The limit of large times formally corresponds to the limit of small amplitude $A$.
The requirement that Eq.~(\ref{eq:anz}) solves Eq.~(\ref{eq:me}) up to terms of order $A^1$ leads to the following
non-linear eigenvalue problem:
\bea
\zeta+\gamma\,(M+3)+\gamma\,\sum_{m=2}^M e^{-\sum_{k=2}^m \log\left(1+\frac{\zeta}{\kappa(k)}\right)}=0.
\label{eq:ce}
\eea
Notice that for gelling kernels,
$$\lim_{m\rightarrow \infty}\sum_{k=2}^m \log\left(1+\frac{\zeta}{\kappa(k)}\right)<\infty$$
Therefore, the sum in the left hand side of Eq.~(\ref{eq:ce}) diverges as the first power of $M$:
$$
\lim_{M\rightarrow \infty} \frac{1}{M} \sum_{m=2}^M e^{-\sum_{k=2}^m \log\left(1+\frac{\zeta}{\kappa(k)}\right)}=
e^{-\sum_{k=2}^\infty \log\left(1+\frac{\zeta}{\kappa(k)}\right)}
$$
As we are interested in the limit of large cut-off mass $M$, let us solve the eigenvalue problem, Eq.~ (\ref{eq:ce}),
in the vicinity of $M=\infty$. Bearing in mind that $\gamma\sim M^{-1}$, the leading term of the large 
$M$-expansion of the left hand side of Eq.~(\ref{eq:ce})  gives
\bea
e^{-\log(1+\zeta)-\sum_{k=2}^\infty \log\left(1+\frac{\zeta}{\kappa(k)}\right)}=-1,
\eea
or equivalently,
\bea
\log(1+\zeta)+ \sum_{k=2}^\infty \log\left(1+\frac{\zeta}{\kappa(k)}\right)=\pi i(1 +2 q)\label{eq:cclm}
\eea
where $q \in \mathbf{Z}$.
We need to find the solution to the above equation with the largest real part. 
It turns out that an analytical solution is possible for the set of reaction rates Eq.~(\ref{eq:cc}) in the 
limit $\epsilon \ll 1.$ Let us Taylor expand the 
logarithms in Eq.~(\ref{eq:cclm}) and switch the orders of summation in anticipation of seeking
an asymptotic expansion for $\zeta$. We obtain:
\begin{displaymath}
\sum_{n=1}^\infty (-1)^{n+1}\,S_n\,\zeta^n = \pi i(1 +2 q)
\end{displaymath}
where
\begin{displaymath}
S_n = \frac{1}{n}\,\left(1+\sum_{k=2}^{\infty} \kappa(k)^{-n}\right).
\end{displaymath}
For the reaction rates given by Eq.~(\ref{eq:cc}) with $\epsilon \ll 1$, we have the following
behaviour for the $S_n$:
\begin{eqnarray*}
S_1= &\sim& \frac{1}{\epsilon},\\
S_n= &\sim& 1, ~n=2,3,\ldots.
\end{eqnarray*}
The fact that $S_1\gg S_n$ for small values of $\epsilon$ allows one to find the solution to Eq.~(\ref{eq:cclm}) 
with the largest real part ($q=0$ or $q=-1$) as an asymptotic expansion in $\epsilon$:
\bea
\nonumber \zeta &=&\pm i\left(\pi \frac{1}{S_1}+\pi^3 \frac{S_3}{S_1^4}-2\pi^3\frac{S_2^2}{S_1^5}\right)+
(-\pi^2\frac{S_2}{S_1^3}+\pi^4\frac{S_4}{S_1^5})\\
& &+O(\epsilon^6) \label{eq:mu}.
\eea
We observe that $\mathrm{Re}\,\zeta<0$ and $|\mathrm{Im}\,\zeta|\gg|\mathrm{Re}\,\zeta|$. Therefore the monomer concentration slowly decays to the stationary value in an oscillatory manner with many periods of oscillations per 
inverse decay rate. Accordingly the oscillations' quality factor
is close to $1$ for $\epsilon \ll 1$:
\bea
Q=e^{-2\pi^2 \frac{S_2}{S_1^2}+O(\epsilon^3)}.
\eea
The independence of the quality factor of the cut-off mass $M$ (up to possible logarithmic corrections) is confirmed.
It is also straightforward to check that the physical period of oscillations is $M$-independent: notice that $\mathrm{Re}\, \zeta<0$. Therefore,
at large times the transformation Eq.~(\ref{eq:physt}) reads:
$$
\tau(t)=\tau_0+N_1(\infty)t+O(e^{\mathrm{Re}\, \zeta\,t}).
$$
Substituting this expression into Eq.~(\ref{eq:anz}) we find that in the limit $M\rightarrow \infty$,
\bea
N_1(\tau)=\sqrt{J}(1+Ae^{\zeta \sqrt{J}t}),
\label{eq:per}
\eea
where $\zeta$ is given by Eq.~(\ref{eq:mu}).
Therefore, the period of oscillations does not depend on the cut-off. It scales as the
inverse square root of the flux of monomers:
\bea
T=\frac{2}{\sqrt{J}} S_1 (1+O(\epsilon^3)).
\eea
The period diverges in the limit $\epsilon \rightarrow \infty$, which implies that oscillations
disappear as we increase the locality of the aggregation kernel. These conclusions are in agreement with our 
numerical measurements on the full coagulation equation presented in Sec.~\ref{sec-oscillations}.

\section{Conclusions}
To conclude, the fact that certain aspects of the  regularized Smoluchowski
equation, such as the total density, depend so weakly  on the value of the
cut-off, $M$, used to regularise the system means that instantaneously gelling  
kernels are still potentially reasonable as physical models. Their use to describe
gravitational clustering or differential sedimentation is not in contradiction
with mathematical results which indicate that the density vanishes for all positive
time in the unregularised system provided one is willing to accept non-universal
dependencies on a cut-off as part of the model. We have presented an essentially
complete analysis of the stationary state of the regularised system in the 
presence of a source of monomers when the regularisation is done be removing
all clusters having size larger than the cut-off. In this case, a power
law scaling is observed for large cluster sizes whose exponent is universal
(depending only on the scaling properties of the kernel) but whose prefactor
is strongly cut-off dependent. The scale-to-scale mass balance is of a 
nonlocal character rather than being described by a mass ``cascade" as is the
case for regular gelling and non-gelling kernels.  
Finally, we found that this stationary state is approached in an interesting way
with long-lived oscillatory transient resulting from interaction between
the cut-off and the source. A quantitative analysis of the approach to the steady state was
provided in the simplified case of the addition model which expresses the 
most extreme case of nonlocal mass transfer in which clusters can only grow
by interaction with monomers.

\section*{Acknowledgements}

C.C. would like to acknowledge useful conversations with E. Ben-Naim, 
P. Krapivsky and S. Nazarenko.


%
\end{document}